\documentclass[%
twocolumn,
superscriptaddress,
 amsmath,amssymb,
 aps,
 prx,
]{revtex4-2}

\usepackage{graphicx}
\usepackage{dcolumn}
\usepackage{bm}
\usepackage{hyperref}
\usepackage{physics}
\usepackage{mathtools}
\usepackage{mathrsfs}
\usepackage{siunitx}
\usepackage{lineno}

\usepackage{xcolor}
\definecolor{amber}{rgb}{1,0.49,0}
\definecolor{darkgreen}{rgb}{0,0.55,0}

\DeclareFontFamily{U}{mathx}{\hyphenchar\font45}
\DeclareFontShape{U}{mathx}{m}{n}{<-> mathx10}{}
\DeclareSymbolFont{mathx}{U}{mathx}{m}{n}
\DeclareMathAccent{\widebar}{0}{mathx}{"73}

\usepackage[normalem]{ulem}
\definecolor{tangerine}{rgb}{0.944,0.522,0}
\definecolor{verde}{rgb}{0.,0.6,0}
\definecolor{rosso}{rgb}{0.9,0.0,0.2}
\definecolor{magenta}{rgb}{0.9,0.2,0.9}

\newif\ifhighlight

\newcommand{\highlight}{\highlighttrue}
\highlight
\newcommand{\editor}[2]{%
  \expandafter\newcommand\csname #1note\endcsname[1]{%
    \textcolor{#2}{(\textbf{#1:} ##1)}}%
  \expandafter\newcommand\csname #1\endcsname[1]{%
    \ifhighlight\textcolor{#2}{##1} \else ##1\fi}%
  \expandafter\newcommand\csname #1cancel\endcsname[1]{%
    \ifhighlight\textcolor{#2}{\sout{##1}}\fi}%
  \expandafter\newcommand\csname #1change\endcsname[2]{%
    \ifhighlight\textcolor{#2}{\sout{##1} ##2}\else ##2\fi}%
  \newenvironment{#1text}{\ifhighlight\color{#2}\fi}{\color{black}}
}

\editor{SB}{tangerine}
\editor{PP}{verde}
\editor{resub}{cyan}
\editor{AF}{blue}
\editor{DD}{purple}
\editor{CAVEAT}{red}

\newcommand{\Si}{\mathrm{Si}}
\newcommand{\Ge}{\mathrm{Ge}}

\usepackage{todonotes}
\usepackage{orcidlink}

\usepackage{textcomp}
\def\qe{\textsc{Quantum ESPRESSO}\texttrademark}

\begin{document}

\title{
Effects of colored disorder on the heat conductivity of SiGe alloys from first principles
}

\author{Alfredo Fiorentino\,\orcidlink{0000-0002-3048-5534}}\email{afiorent@sissa.it}
\affiliation{%
 SISSA---Scuola Internazionale Superiore di Studi Avanzati, Trieste
}%
\author{Paolo Pegolo\,\orcidlink{0000-0003-1491-8229}}
\affiliation{%
 Laboratory of Computational Science and Modeling, IMX, École Polytechnique Fédérale de Lausanne,
1015 Lausanne, Switzerland
}

\author{Stefano Baroni\,\orcidlink{0000-0002-3508-6663}} 
\affiliation{%
 SISSA---Scuola Internazionale Superiore di Studi Avanzati, Trieste
}%
\affiliation{%
 CNR-IOM---Istituto Officina Materiali, DEMOCRITOS SISSA unit, Trieste
}%

\author{Davide Donadio\,\orcidlink{0000-0002-2150-4182}}
\affiliation{%
 Department of Chemistry, University of California, Davis, Davis, California 95616, United States}%

\date{\today}

\begin{abstract}

Semiconducting alloys, in particular SiGe, have been employed for several decades as high-temperature thermoelectric materials. Devising strategies to reduce their thermal conductivity may provide a substantial improvement in their thermoelectric performance also at lower temperatures. We have carried out an {\sl ab initio} investigation of the thermal conductivity of SiGe alloys with random and spatially correlated mass disorder employing the Quasi-Harmonic Green-Kubo (QHGK) theory with force constants computed by density functional theory. Leveraging QHGK and the hydrodynamic extrapolation to achieve size convergence, we obtained a detailed understanding of lattice heat conduction in SiGe and demonstrated that colored disorder suppresses thermal transport across the acoustic vibrational spectrum, leading to up to a 4-fold enhancement in the intrinsic thermoelectric figure of merit.    

\end{abstract}

\maketitle

\section{Introduction}

Thermoelectric (TE) devices are instrumental in managing heat and converting otherwise wasted thermal energy into useful electrical power. The efficiency of TE materials is quantified by the dimensionless figure of merit:
\begin{align*}
    ZT=\frac{\sigma S^2 T}{\kappa},
\end{align*}
where $\sigma$ is the electrical conductivity, $S$ the Seebeck coefficient, $T$ the temperature, and $\kappa=\kappa^{\mathrm{el}}+\kappa^\mathrm{l}$ encompasses the thermal conductivity, computed as the sum of electronic (el) and lattice (l) contributions. While maximizing the figure of merit by manipulating electronic properties $(\sigma, S, \kappa^{\mathrm{el}})$ is often complicated and limited by their interconnection, a successful strategy involves reducing the lattice thermal conductivity without significantly altering the other factors~\cite{snyder_complex_2008,beretta_thermoelectrics_2019,pecunia_roadmap_2023}. 
For silicon-based TE devices, a viable way to achieve this goal is through alloying with germanium~\cite{steele1958thermal,dismukes1964thermal}. Indeed, through doping, silicon can become a TE device with a high power factor ($\sigma S^2 \sim 22\mathrm{\mu W cm^{-1}K^2}$ at room temperature~\cite{stranz_thermoelectric_2013}), and $\sigma S^2$ has a weak dependence on alloying with Ge~\cite{steele1958thermal,hahn2021intrinsic}. 
However, its performance is hindered by a high lattice thermal conductivity ($\approx 148$-$156\,\mathrm{Wm^{-1}K^{-1}}$ at room temperature\cite{kremer2004thermal}). Through alloying with a chemically similar element such as germanium, the lattice thermal conductivity of bulk pure silicon can be reduced by at least an order of magnitude~\cite{garg2011role, he_lattice_2012, hahn2021engineering, lorenzi_phonon_2018, xiong_native_2017} without altering significantly the electronic transport coefficients~\cite{de1991structure,marzari1994structure,hahn2021intrinsic}. SiGe alloys have been used since the 1960s in the high-temperature regime for applications like powering space probes with typical $ZT \approx 1$ at the operating temperature of $T\approx 1100\,$K~\cite{wang_enhanced_2008}. 
Further development in the field would allow $\Si_{1-x}\Ge_x$ alloys to compete, at room temperature, with the class of TE devices presently employed, often composed of rare and/or toxic materials~\cite{donadio2019advances}. Such a goal has been pursued through the years through alloying and structuring at the nano and mesoscale~\cite{wang_enhanced_2008, bathula2012enhanced, savic2013dimensionality,ferrando-villalba_tailoring_2015,ferrando-villalba_beating_2020}. Both theoretical~\cite{thebaud2023breaking} and experimental~\cite{chaney2021tuneable} work have recently suggested the use of spatially correlated disorder to strongly damp acoustic vibrations.

The goal of this work is to study from first principles the lattice thermal conductivity of $\Si_{1-x}\Ge_{x}$ alloys, both for the standard uncorrelated (\emph{white}) and spatially correlated (\emph{colored}) compositional disorder. The theoretical foundation of our calculations is the recently developed Quasi-Harmonic Green-Kubo theory (QHGK)~\cite{isaeva2019modeling}. This theory, along with a different formulation based on the Wigner Boltzmann Transport Equation~\cite{simoncelli2019unified}, offers a unified approach to the thermal conductivity of solid insulators in the weakly anharmonic limit, both for crystals and disordered systems like glasses or alloys. Therefore, both these theories fully account for the effects of disorder on the properties of alloys, beyond the standard perturbative approach~\cite{garg2011role,hahn2021engineering}. However, the computational cost of QHGK on disordered systems scales poorly with the system size, severely hindering our ability to reach the bulk infinite-size limit. This challenge can be overcome using the \emph{hydrodynamic extrapolation}~\cite{fiorentino2023hydrodynamic}, which allows for a speedy and accurate evaluation of the low-frequency sound-wave contribution to the heat conductivity that is the culprit for poor size convergence~\cite{fiorentino2023unearthing}. The hydrodynamic extrapolation relies on the computation of the Vibrational Dynamical Structure Factor (VDSF) and its connection to hydrodynamic equations. Furthermore, the VDSF can be efficiently computed~\cite{fiorentino2023unearthing} on systems of more than a hundred thousand atoms, which exceeds by an order of magnitude the largest disordered systems directly treated using QHGK in existing literature~\cite{fiorentino2023hydrodynamic,fiorentino2023unearthing}. This aspect is fundamental for studying spatially correlated systems, particularly when the correlation length is significantly larger than the average interatomic distance.

The article is organized as follows: after a quick review of the QHGK theory and of the hydrodynamic extrapolation, we benchmark our method for uncorrelated alloys against experimental measurements on Raman spectroscopy and thermal conductivity. We then examine the effect of different instances of spatially correlated disorder on the harmonic scattering and consequently on the thermal conductivity. Finally, we present our conclusions.

\section{Theory}

\subsection{Quasi-Harmonic Green-Kubo approximation} \label{subsec:QHGK}

The QHGK thermal conductivity is obtained solving the quantum GK formula~\cite{green1952markoff,kubo1957statstical1,kubo1957statistical2} in the quasi-harmonic regime~\cite{isaeva2019modeling, fiorentino2022green, simoncelli2019unified, caldarelli2022manybody}:
\begin{align}
    \label{eq:k_markovian}
    \kappa^{\alpha\beta} = \frac{1}{V}\sum_{\mu \mu'} C_{\mu \mu'} v_{\mu\mu'}^\alpha v_{\mu' \mu}^\beta \tau_{\mu \mu'},
\end{align}
where $\alpha,\beta$ are cartesian indices, and
\begin{align}
    C_{\mu \mu'} = \frac{\hbar^2 \omega_{\mu}\omega_{\mu' }}{T}\frac{n_{\mu}-n_{\mu' }}{\hbar(\omega_{\mu' }-\omega_{\mu})}\\
        \tau_{\mu \mu'} = \frac{\gamma_{\mu}+\gamma_{\mu' }}{(\omega_{\mu}-\omega_{\mu' })^2+(\gamma_{\mu}+\gamma_{\mu' })^2}
\end{align}
are respectively the generalized two-mode heat capacity and two-mode lifetime; $\omega_{\mu' }$ and $\gamma_{\mu' }$  are the normal-mode angular frequency and anharmonic linewidth; ${n_{\mu}=(e^{\hbar\omega_{\mu}/k_{\mathrm B}T}-1)^{-1}}$ the Bose-Einstein distribution, and $v_{\mu \mu'}$ is a generalized two-mode velocity matrix~\cite{isaeva2019modeling}. For crystals, it is convenient to use the Bloch states as a basis for the eigenvectors and therefore to label phonons as ${\mu = (\mathbf{q}, s)}$, where $\mathbf{q}$ and $s$ are the crystal momentum and band index, respectively. In this basis, the generalized velocity matrix is block-diagonal with respect to crystal momentum, $v_{\mathbf{q}\mathbf{q}'ss'} = \delta_{\mathbf{q}\mathbf{q}'} v_{\mathbf{q}ss'}$, and its diagonal elements in the band indices are the usual group velocities, $v_{\mathbf{q}ss}^\alpha = \frac{\partial \omega_{\mathbf{q}s}}{\partial q_\alpha}$.
 
Labeling normal modes of disordered systems with the phonon crystal momentum would seem unnatural and to some extent misleading, due to the lack of periodicity. Notwithstanding, this practice is sometimes justified by the common use of periodic boundary conditions (PBCs) and adopted to enhance the sampling of vibrational modes. It has been shown that a full sampling of the (mini) Brillouin zone (BZ) corresponding to the simulation supercell of a glass model introduces an unphysical order at scales larger than the simulation cell, possibly leading to uncontrolled errors in the computed thermal conductivity~\cite{moon2018propagating, fiorentino2023hydrodynamic,  fiorentino2023unearthing}.
Therefore, throughout this work the normal modes of disordered systems are always computed at the BZ center ($\Gamma$ point) and labeled by the discrete ``band'' indices, $\mu=1,\dots,3N_\mathrm{atoms}$. 

The QHGK formula is derived under the single-mode relaxation time approximation (RTA). In a nutshell, this expression entails two different yet related approximations. Firstly, \emph{single-mode} means that any vertex corrections to 4-point correlation functions, in the many-body parlance, is neglected~\cite{caldarelli2022manybody,fiorentino2022green}. Secondly, the RTA consists in assuming the following \emph{ansatz} for the single-body \emph{greater} (and similarly for the \emph{lesser}) Green's function: ${g_\mu=-i\langle\hat a_\mu^\dag(t)\hat a_\mu\rangle\approx -i(n_\mu+1)e^{i\omega_\mu t-\gamma_\mu|t|}}$, where $\hat a^\dag$/$\hat a$ are the creation/annihilation operators, and ${\gamma_\mu\ll \omega_\mu}$. This corresponds to modeling the interaction of each individual mode out of equilibrium with a bath of phonons featuring equilibrium populations. For the sake of simplicity, we omit the Cartesian indices in the rest of the paper. Unless otherwise specified, we indicate as $\kappa=\frac{1}{3}\sum_\alpha \kappa^{\alpha\alpha}$, the isotropic average of Eq.~\ref{eq:k_markovian}.

\subsection{ Electronic and vibrational Virtual Crystal Approximations for SiGe alloys }\label{subsec:MassApprox}

Lattice dynamical methods like QHGK require the knowledge of the Born-Oppenheimer (BO) potential energy surface close to mechanical equilibrium, which is entirely determined by the electronic ground state at fixed atomic positions and by the atomic masses. Indeed, the dynamical matrix element between two atoms $I$ and $J$ is defined as:
\begin{align*}
    D_{IJ}=\frac{1}{\sqrt{M_{I}M_{J}}}\frac{\partial^2 U}{\partial R_{I}R_{J}}
\end{align*}
where $U$ is the BO potential energy, and $R_I$ and $M_I$ are the equilibrium position and mass of the $I$th atom, respectively, and Cartesian indices have been omitted for notational simplicity.

For an alloy, both the BO potential and the distribution of masses contribute to the disorder, which we will refer to as \textit{chemical} and \textit{mass} disorder, respectively. However, for chemically similar atoms like Si and Ge, it is known~\cite{de1991structure,marzari1994structure} that the BO potential can be accurately described by the (electronic) Virtual Crystal Approximation (eVCA). In essence, for any given concentration of $\Si_{1-x}Ge_{x}$, we consider the electronic properties of a ``virtual'' crystal where both atomic species are replaced by a fictitious one whose (pseudo)potential is the linear interpolation of the actual species' potential. Under this assumption, the dynamical matrix becomes: 
\begin{align*}
    \bar D_{IJ}^\mathrm{e}(x)=\frac{1}{\sqrt{M_{I}M_{J}}}\frac{\partial^2 U_{x}}{\partial R_{I}R_{J}} 
\end{align*}
where $U_x$ and $R$ is the ($x$-dependent) BO potential energy, while the masses are still distributed at random (for details, see Appendix~\ref{app:generation disorder}). We refer to this case as the mass-disordered alloy, where mass disorder is fully accounted for, while chemical disorder is neglected. If we also disregard the mass disorder, we obtain the vibrational Virtual Crystal Approximation (vVCA). In this case,
\begin{align*}
    \bar D_{IJ}^\mathrm{v}(x)=\frac{1}{M(x)}\frac{\partial^2 U_{x}}{\partial R_{I} \partial R_{J}}
\end{align*}
where $M(x) = xM_\mathrm{Ge} + (1-x)M_\mathrm{Si}$. While being a much cruder approximation\cite{de1991structure}, the vVCA can still provide valuable insights into the vibrational properties of the system in the acoustic region, where neighboring Si and Ge atoms vibrate in phase, and it is often the starting point for perturbative treatments of mass disorder~\cite{garg2011role,hahn2021engineering,hahn2021intrinsic}. 

The eVCA is assumed to be valid throughout this paper and it plays a fundamental role in enabling the computation of second- and third-order Interatomic Force Constants (IFCs) using \emph{ab initio} methods such as Density Functional Perturbation Theory (DFPT)~\cite{Baroni1987a,*Giannozzi1991a,*baroni2001phonons}. By neglecting chemical disorder, the IFCs can be computed efficiently for the virtual crystal, fully leveraging the benefits of periodicity.

\begin{figure*}[t]
    \centering
    \includegraphics[width=1.8\columnwidth]{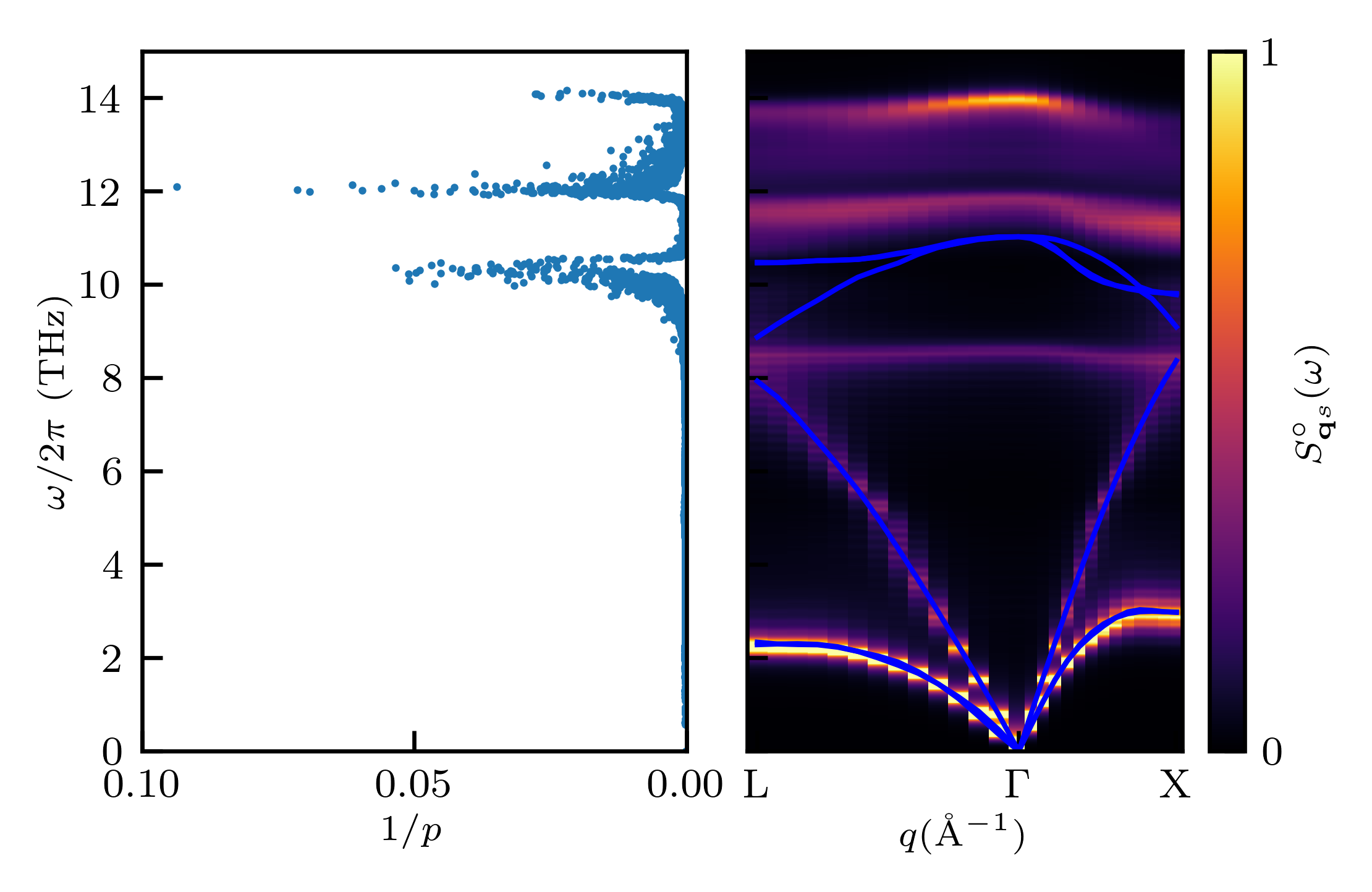}
    \caption{Left panel: Inverse participation ratio of the normal modes of a disordered Si-Ge alloy containing $\approx 12000$ atoms at mixing ratio of $x=0.5$. Right panel: {Spectral function $S_{\mathbf{q}s}^\circ(\omega)$ (Eq. \ref{eq:harmonic_VDSF})} of the same system shown as a colormap, along with the phonon band dispersion (blue) of a virtual crystal of size equivalent to $\approx 45000$ atoms at $x=0.5$ overlaid. The spectral function is computed in the harmonic approximation with an artificial smearing $\eta=0.5\mathrm{ ~rad ~ps^{-1}}$ and {its intensity is expressed in arbitrary units.} 
    Spikes in $1/p$ are observed in the optical frequency range, corresponding to regions where the differences between the vVCA dispersion and the VDSF are more pronounced. 
    }
    \label{fig:dispersion&ipr}
\end{figure*}

\subsection{Scattering by mass disorder}

For any concentration $x$ ranging from 0 to 1, the $\Si_{1-x}\Ge_x$ alloy in the eVCA must be modeled as a disordered medium with respect to lattice vibrations, due to the disruptive impact of mass disorder on the periodicity of the dynamical matrix. As in the case of other disordered solids, such as amorphous ones, the lack of periodicity poses a significant challenge in achieving size convergence through a direct approach. One major computational bottleneck is the diagonalization of the dynamical matrix, which scales poorly (cubically) with the number of atoms. The finite size particularly impacts low frequencies, since the minimum nonzero frequency scales as $\omega_{\mathrm{min}}\sim c/L$, where $L$ is the edge length of a cubic simulation cell, and $c$ the speed of sound. While treating hundreds of thousands of atoms in reciprocal space is routine for crystals, by exploiting periodicity through the Bloch theorem, it is computationally unfeasible to treat the same number of atoms in disordered media, despite this being sometimes required to achieve convergence~\cite{fiorentino2023hydrodynamic}.

To address this challenge, various techniques have been explored~\cite{garg2011role,thebaud2020success,fiorentino2023hydrodynamic}. For the uncorrelated case, we employ and compare two distinct approaches: the standard perturbative solution~\cite{garg2011role} and the recently developed \emph{hydrodynamic extrapolation}~\cite{fiorentino2023hydrodynamic} which fully accounts for the harmonic disorder. For the spatially correlated case, we rely only on hydrodynamic extrapolation, since a perturbative treatment in the presence of spatial correlations would require further theoretical and numerical effort.

The standard approach for an uncorrelated alloy involves a perturbative treatment of disorder~\cite{garg2011role}. The phonon frequencies and velocities of the virtual crystal are used in the Boltzmann Transport Equation within the single-mode Relaxation Time Approximation (BTE-RTA):
\begin{align}\label{eq:bte-rta}
    \kappa^{\mathrm{BTE\text{-}RTA}}=\frac{1}{3V}\sum_{\mathbf{q} s} C_{\mathbf{q}s} v_{\mathbf{q} s}v_{\mathbf{q}s} \tau_{\mathbf q s},
\end{align}
where the effect of disorder scattering is accounted for in the lifetimes through Matthiessen's rule~\cite{matthiessen1862influence}: ${\tau_{\mathbf q s}^{-1}=(\tau_{\mathbf q s}^{\mathrm{anh}})^{-1}+(\tau_{\mathbf q s}^{\mathrm{iso}})^{-1}}$.  The mass disorder linewidth $\Gamma_{\mathbf q s}^{\mathrm{iso}}$ can be computed perturbatively with Fermi's Golden Rule (FGR), yielding to Tamura's formula for perturbative isotopic scattering~\cite{tamura1983isotope}, which for a crystal with only one chemical element reads:
\begin{align}\label{eq:tamura_scattering}
    \frac{1}{\tau_{\mathbf q s}^{\mathrm{iso}}}&=\frac{\pi}{2}{g_2}\omega_{\mathbf q s}^2\rho(\omega_{\mathbf q s})\\
    &=2\Gamma_{\mathbf q s}^{\mathrm{iso}}.
\end{align}
Here, $\rho$ is the vibrational density of states normalized to unity, and $g_2$ is a measure of mass variance, given by:
\begin{align*}
    {g_2}=\sum_i^{n_\mathrm{types}}f_i\left(1-\frac{m_i}{\overline{m}}\right)^2,
\end{align*}
involving mass ($m_i$) and concentration ($f_i$) of the $i$th among $n_\mathrm{types}$ species, and their average mass, $\overline{m}$.

While the perturbative approach using the vVCA normal modes is effective for acoustic bands, it fails in the optical frequency range when alloying effects are sufficiently large. 

This breakdown is demonstrated in Fig.~\ref{fig:dispersion&ipr} using 
the spectral function of the virtual crystal eigenvectors in the harmonic approximation
\begin{align}\label{eq:harmonic_VDSF}
    S_{\mathbf{q}s}^\circ(\omega) = \lim_{\eta\to 0} \sum_{\mu} \frac{1}{\pi} \frac{\eta}{\eta^2+(\omega-\omega_\mu)^2}|\langle \mu| \mathbf{q}s\rangle|^2,
\end{align}
where $\langle \mu| \mathbf{q}s\rangle$ is the scalar product between the vVCA eigenvector and the eVCA one, and the Inverse Participation Ratio (IPR)~\cite{feldman1993thermal},
\begin{equation} \label{eq:IPR}
    \frac{1}{p_\mu} =\frac{ \sum_{I}[\sum_{\alpha}(e_{I\alpha}^\mu)^2]^2}{\sum_{I\alpha}(e_{I\alpha}^\mu)^2}
\end{equation}
where $ e_{I\alpha}^\mu$ is  the component of the $\mu$-th vibrational eigenvector on the $I$-th atom in Cartesian direction $\alpha$. As observed in the spectral function, the vVCA cannot predict the splitting of the optical bands into Si-Si, Si-Ge, and Ge-Ge-related modes\cite{de1992effects}. Moreover, a significative difference between eVCA and vVCA is indicated by the IPR, which is a measure of localization and ranges from $1/N_\mathrm{atoms}\approx 0$ (fully delocalized modes) to $1$ (fully localized). The IPR shows that optical modes undergo substantial localization that cannot be captured in the vVCA approach, in which the normal modes are crystalline, hence fully delocalized.

The failure of the crystal picture at higher frequencies has been observed in various alloys~\cite{de1992effects,thebaud2020success,thebaud2022perturbation}, resembling the behavior of vibrational modes in amorphous solids. In glasses, low frequencies are populated with acoustic-like propagating modes, known as \textit{propagons}. As the energy increases, vibrations lose their propagating character and cannot be assigned a group velocity. These modes may be delocalized, called \textit{diffusons}, or localized, \textit{locons}, and contribute to the diffusive component of the thermal conductivity in glasses~\cite{allen1999diffusons}. 

Driven by the analogy with the theory of heat transport in glasses, we employ the hydrodynamic extrapolation technique to significantly accelerate size convergence~\cite{fiorentino2023hydrodynamic}. Essentially, this method involves separating the low-energy propagons' contribution from the diffusive one and using an acoustic wave-like basis to compute the former. Leveraging arguments from the hydrodynamics of solids, the propagons' contribution can be extrapolated to the bulk system~\cite{griffin1968brillouin, fiorentino2023hydrodynamic}. For isotropic thermal conductivity, the hydrodynamic extrapolation is expressed as~\cite{fiorentino2023hydrodynamic}:
\begin{align}
    &\kappa_\mathrm{hydro}=\kappa_\mathrm{P}+\kappa_\mathrm{D} \label{eq:kappa hydro}\\
    &\kappa_\mathrm{P} = \frac{1}{3V}\sum_{\mathbf{q} b} C_{\mathbf{q}b} |v_{\mathbf{q}b}|^2 \frac{1}{2\Gamma_{\mathbf{q}b}}\Theta(\omega_\mathrm{P} - \omega_{\mathbf q b}) \label{eq:kappa P}\\
    &\kappa_\mathrm{D}= \frac{1}{3V} \sum_{\mu \mu'} \Theta(\omega_{\mu}-\omega_\mathrm{P})\Theta(\omega_{\mu'}-\omega_\mathrm{P}) C_{\mu \mu'} |v_{\mu \mu'}|^2 \tau_{\mu \mu'}. \label{eq:kappa D}
\end{align}
Here, the index $b\in\{ T_1,T_2,L\}$ indicates the transverse or longitudinal polarization of the acoustic branches, and $\Theta$ is the Heaviside function that separates the propagonic and non-propagonic contributions~\cite{fiorentino2023hydrodynamic,fiorentino2023unearthing}.
The propagon contribution is then computed on a dense grid of wavevectors, commonly referred to as a $q$-mesh, and it can be easily extrapolated to the bulk limit, resulting in the inclusion of a Debye-like term in the thermal conductivity~\cite{fiorentino2023hydrodynamic,fiorentino2023unearthing}.

The diffusive thermal conductivity, $\kappa_\mathrm{D}$, which is to a large extent size-insensitive, is computed using the normal modes of a disordered finite cell, while the propagon one, $\kappa_\mathrm{P}$, involves the acoustic branches below a certain frequency of the size-converged virtual crystal. The transition between the two regimes is not sharp, allowing some freedom in the choice of $\omega_\mathrm{P}$. However, as discussed in Ref.~\onlinecite{fiorentino2023hydrodynamic} and in Appendix~\ref{app:hydro_ext}, as long as $\omega_\mathrm{P} > 
\omega_{\mathrm{min}}$ is in a region where the acoustic branches are well-defined and distinct by polarization, then the thermal conductivity depends only very weakly on $\omega_\mathrm{P}$. The propagons linewidth, $\Gamma_{\mathbf q b}$, accounts for both harmonic mass-disorder (non-perturbatively, as detailed below) and perturbative third-order anharmonic scattering affecting the vVCA vibrations, unlike the purely anharmonic linewidths of the normal modes, $\gamma_\mu$.  

Estimating $\Gamma_{\mathbf q b}$ involves computing the VDSF of acoustic phonons, whose straightforward generalization to include perturbative anharmonic effects is~\cite{fiorentino2023hydrodynamic}:
\begin{align*}
S_{\mathbf q b}(\omega)=\sum_{\mu}\frac{1}{\pi}\frac{\gamma_\mu}{\gamma_\mu^2+(\omega-\omega_\mu)^2}|\langle \mu| \mathbf{q}b\rangle|^2.
\end{align*}
Then, the linewidths $\Gamma_{\mathbf{q}b}$ are evaluated by a Lorentzian fit~\cite{feldman1999numerical,fiorentino2023hydrodynamic}:
\begin{align}
    S_{\mathbf q b}(\omega)\approx \frac{A_{\mathbf q b}}{\pi}\frac{\Gamma_{\mathbf q b}}{(\omega-\omega_{\mathbf q b})^2 +\Gamma_{\mathbf q b}^2},
\end{align}
where $A_{\mathbf q b}$ is a normalization factor. Analogously, the linewidth $\Gamma_{\mathbf{q}b}^\circ$ due to harmonic mass-disorder alone is obtained by a Lorentzian fit of $S_{\mathbf{q}b}^\circ$.

As shown in Ref.~\onlinecite{fiorentino2023unearthing}, the VDSF and, more generally, the diagonal elements of the vibrational Green's functions can be efficiently computed using the Haydock recursion method~\cite{haydock1980recursive,vast2000effects}, leveraging the iterative Lanczos technique. This procedure avoids diagonalization and scales as $\mathcal O(P)$, $P$ being the number of non-vanishing elements of the dynamical matrix: $P\propto N_\mathrm{atoms}^2$ for a dense dynamical matrix, $P\propto N_\mathrm{atoms}$ if the dynamical matrix is sparse, as in the case of short-range interactions. It is worth mentioning that in the eVCA long-range effects could also be treated efficiently by using a different basis for the Haydock method. As shown in Ref.~\onlinecite{vast2000effects}, nearly linear scaling can be achieved by working simultaneously in reciprocal and real space, rather than relying exclusively on a real-space implementation as done in this work.
This speed-up enables the study of the VDSF of disordered systems with hundreds of thousands of atoms~\cite{fiorentino2023unearthing}. 

To summarize, the diffusive contribution to thermal conductivity is obtained from disordered samples of manageable size, while $\kappa_\mathrm{P}$ is efficiently computed using the acoustic dispersion of the virtual crystal and the linewidths extracted from the VDSF, whose calculation is accelerated by the Haydock method~\cite{haydock1980recursive}. Although $\kappa_\mathrm{P}$ (Eq.~\ref{eq:kappa D}) and Eq.~\ref{eq:bte-rta} seem similar when restricted to the acoustic branches, it is crucial to note that the VDSF, and therefore $\Gamma_{\mathbf{q}b}$, fully account for mass disorder, beyond the perturbative level of Eq.~\ref{eq:tamura_scattering}.  

We finally mention that Ref.~\onlinecite{thebaud2022perturbation} recently proposed a method to solve the Green-Kubo equation at the quasi-harmonic level, as in QHGK, which substitutes the hurdle of diagonalizing massive dynamical matrices with the repeated use of stochastic traces. While this method cannot substantially change the performance of computing $\kappa_\mathrm{P}$ with the VDSF, it could further reduce the computational cost of the diffusive part.

\section{Results}

The virtual crystal $\Si_{1-x}\Ge_x$ diamond structures are simulated, including the calculation of second- and third-order interatomic force constants, using the Density Functional Theory codes in the \qe~\cite{giannozzi2009quantum,*giannozzi2017advanced,*giannozzi2020quantum} (QE) distribution, and \texttt{D3Q}~\cite{paulatto2013anharmonic}, which implements third-order DFPT. Lattice dynamics calculations are performed with $\kappa$ALD$o$~\cite{barbalinardo2020efficient}. Computational details are given in Appendix~\ref{app:computational}. 

\begin{figure}[htb!]
    \includegraphics[width=\columnwidth]{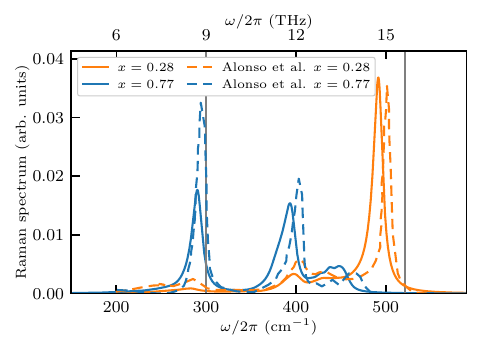}
    \caption{Unpolarized Raman spectroscopy at room temperature for different Germanium concentration $x$. Dashed lines:  experimental data from Ref.~\onlinecite{alonso1989raman}. Continuous lines: numerical Raman spectrum computed as explained in the main text with smearing equal to the experimental resolution $\eta=5\,\mathrm{cm^{-1}}$. Vertical lines represent, respectively, the experimental~\cite{alonso1989raman} pure Germanium and Silicon bulk degenerate optic frequencies at $q=(0,0,0)$, $\omega_{\Ge}$ and $\omega_{\Si}$.  To take into account the small discrepancies between the DFT frequencies and the experimental ones ($\leq 2\%$) of the pure systems $\Delta_{\Ge/\Si}=\omega_{\Ge/\Si}^{exp}-\omega_{\Ge/\Si}^{DFT}$, the numerical spectra have been shifted by $\Delta(\omega)=\frac{\omega -\omega_{\Ge}}{\omega_{\Si} -\omega_{\Ge}}\Delta_{\Si}+\Delta_{\Ge}$.}
    \label{fig:raman}
\end{figure}

\subsection{Raman spectroscopy}


We test the reliability of our predictions by computing the unpolarized nonresonant Stokes Raman spectrum of $\Si_{1-x}\Ge_x$ alloys using the Placzek approximation~\cite{bresch1986phonons}. Following 
Ref.~\onlinecite{vast2000effects}, the computation of the polarized Raman spectrum closely resembles that of the VDSF, as explained in detail in Appendix~\ref{app:raman}. To compare the experimental and numerical results, the spectrum is computed with a smearing width of $\eta=5\,\mathrm{cm^{-1}}$, matching the experimental resolution~\cite{alonso1989raman}. The vVCA anharmonic linewidths at room temperature are neglected as they are an order of magnitude smaller than the experimental resolution in the selected frequency range, and the line broadening is dominated by mass disorder, which is explicitly accounted for in the VDSF calculation.

The experimental phenomenology of $\Si_{1-x}\Ge_{x}$ alloys has been extensively studied~\cite{alonso1989raman,mooney1993raman,rouchon2014germanium}. The Raman spectra of pure $\Si$ and $\Ge$ systems exhibit peaks at the degenerate optical frequencies at the $\Gamma$ point of the BZ, located at $521\,\mathrm{cm^{-1}}$ and $300\,\mathrm{cm^{-1}}$, respectively. At intermediate concentrations, $\Si$, $\Si$/$\Ge$, and $\Ge$ peaks are observed near their characteristic frequencies, with broadening primarily due to harmonic disorder. Additionally, minor peaks between $400\,\mathrm{cm^{-1}}$ and $500\,\mathrm{cm^{-1}}$ are commonly observed in experiments. Our calculations accurately predict the observed positions of both the major and minor peaks. However, the experimental $\Ge$ peak appears sharper than the simulated one. This discrepancy could be attributed to several factors, including possible frequency-dependent variations in experimental resolution\cite{schrader1990nir} and the chemical bonding differences between $\Si$ and $\Ge$ sites, which are not accounted for under the eVCA assumptions. It is worth mentioning that the vVCA approach would predict only one peak as the pure systems, with a concentration-dependent position, again proving a qualitative difference between the vVCA and eVCA in describing vibrations in the optical frequency range \cite{de1992effects}. 

\subsection{Thermal conductivity of $\Si_{1-x}\Ge_{x}$ alloys}
\begin{figure}[htb!]
    \includegraphics[width=\columnwidth]{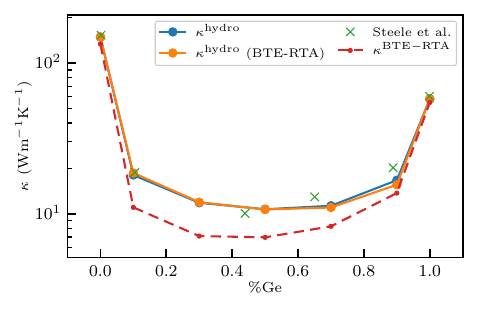}
    \includegraphics[width=\columnwidth]{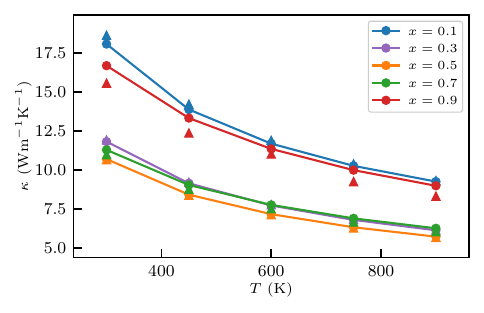}
    \caption{ Upper panel: bulk thermal conductivity at room temperature as a function of Germanium concentration, computed with both the BTE-RTA and QHGK methods and compared to experimental results\cite{steele1958thermal}. The BTE-RTA results are also shown for a dense $q$-mesh $[28,28,28]$ (dashed line) to stress the importance of finite-size effects. Lower panel: bulk thermal conductivity as a function of temperature and concentration. The continuous lines are QHGK results, while the triangles of the same color are the corresponding BTE-RTA ones.}
    \label{fig:tk_alloy}
\end{figure}

We compute the thermal conductivity of uncorrelated alloys, comparing our method to the BTE-RTA, which has successfully reproduced experimental  measurements~\cite{garg2011role,hahn2021engineering,hahn2021intrinsic}. 
As observed in Ref.~\onlinecite{hahn2021engineering} and Fig.~\ref{fig:tk_alloy}, even with a typically dense $q$-mesh with a spacing between BZ points of 0.07~\AA$^{-1}$, i.e. a $30 \times 30 \times 30$ mesh, finite-size effects can reduce the bulk thermal conductivity at room temperature by up to $40\%$. 
We cope with finite-size effects through the hydrodynamic extrapolation, as explained in Ref.~\onlinecite{fiorentino2023hydrodynamic}. 
In the BTE-RTA method, the hydrodynamic extrapolation simply consists of including the Debye contribution on top of the results obtained with a dense $q$-mesh.
To perform the extrapolation, we fit the scattering for vanishing frequencies as ${\Gamma(\omega)=a(T)\omega^2 + b \omega^4}$~\cite{fiorentino2023hydrodynamic,larkin2014thermal}, where the temperature-dependent quadratic term is due to the anharmonic linewidths and the quartic term is due to disorder. In addition to the mass disorder due to alloying, we include isotopic scattering computed with Eq.~\ref{eq:tamura_scattering} for both methods. Unsurprisingly, isotopic effects are impactful only for the pure systems, as the scattering due to alloying is on average two to three orders of magnitudes larger for the intermediate concentrations. 

The bulk thermal conductivity is shown in Fig.~\ref{fig:tk_alloy} as a function of the concentration of Ge (upper panel) and of the temperature (lower panel). At fixed temperature, both methods exhibit a distinctive U-shape, typical of both crystalline~\cite{abeles1962thermal,garg2011role} and amorphous~\cite{lundgren2021mode,pegolo2024thermal} alloys, with a minimum around $x\approx 0.5$. Notably, the results from both methods are nearly equivalent, with a relative difference within $10\%$. 
{Both methods reproduce well the experimental data at room temperature~\cite{steele1958thermal}. However, there is significant variance in experimental results, e.g., for $T=300\,$K and $x=0.5$, $\kappa$ can vary between $6$ and $11\,\mathrm{Wm^{-1}K^{-1}}$\cite{steele1958thermal,abeles1962thermal,dismukes1964thermal}. 
This variance suggests some caution in the comparison with experiments and underscores the importance of accurately characterizing the disorder. In fact, according to Ref.~\onlinecite{steele1958thermal}, the process employed to grow the alloy and the resulting disorder is the most likely cause of such variance.} 

\begin{figure}[htb!]
    \includegraphics[width=\columnwidth]{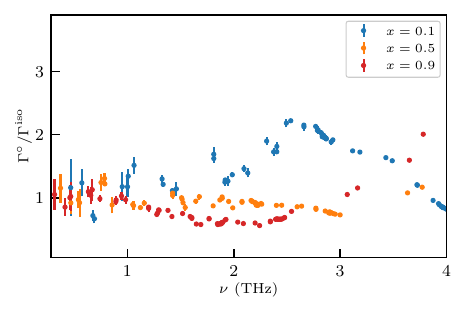}
    \caption{Ratio between the mass disorder linewidth computed from the non-perturbative and perturbative approaches, respectively Eq.~\ref{eq:harmonic_VDSF}-\ref{eq:tamura_scattering}. The error bars are the standard deviation over $4$ samples of $N\approx 45000$ atoms.}
    \label{fig:fgr_comparison}
\end{figure}

The success of the perturbative approach in reproducing the QHGK result, despite overlooking the diffusive contribution and misrepresenting the optical phonons dispersion, is attributed to the dominant role of acoustic phonons, for which a perturbative treatment is reasonable. Indeed, according to Ref.~\onlinecite{garg2011role}, at room temperature and $x=0.5$, phonons with a frequency below $2\,\mathrm{THz}$ contribute  $88\%$ of the total thermal conductivity. 
The lack of a diffusive interband contribution in the BTE-RTA  would seem to imply that this method poses a lower bound to the QHGK results if harmonic scattering were exactly accounted for (see the lower panel of Fig.~\ref{fig:kappa_cumulative}). Instead, $\kappa^{\mathrm{QHGK}} < \kappa^{\mathrm{BTE\text{-}RTA}}$ at low $\Ge$ concentrations, highlighting the importance of a non-perturbative account of mass-disorder scattering. 
As shown in Fig.~\ref{fig:fgr_comparison}, and in agreement with the molecular dynamics results of Ref.~\onlinecite{larkin2013predicting}, even for acoustic modes below $3\,\mathrm{THz}$, linewidths from exact mass-disorder scattering can significantly differ, both positively and negatively, from the perturbative treatment of Eq.~\ref{eq:tamura_scattering}~\cite{tamura1983isotope}. However, it is somewhat surprising that such corrections to Tamura's formula are more pronounced for dilute concentrations ($x=0.1, 0.9$) than for $x=0.5$. A rationale can be found in the prefactor $-g_3 = -\sum_i^{n_{\mathrm{types}}} f_i \left(1 - \frac{m_i}{\overline{m}}\right)^3$ in the third order correction to Eq.~\ref{eq:tamura_scattering}~\cite{tamura1983isotope}, where for $x=0.1,0.5,0.9$, the values for $-g_3$ are approximately $0.2, 0, -0.02$, respectively, which agree with the observations in Fig.~\ref{fig:fgr_comparison}.

\subsection{Colored disorder}

\begin{figure}[htb!]
    \includegraphics[width=\columnwidth]{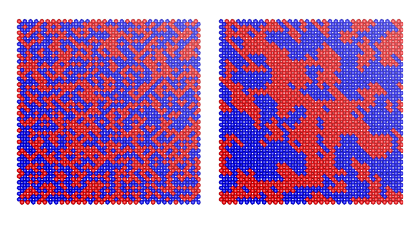}
    \caption{Section of a $\Si_{0.5}\Ge_{0.5}$ alloy model with $\approx 65000$ atoms with uncorrelated (left) and correlated (right) mass disorder. The red and blue dots are respectively $\Si$ and $\Ge$ atoms. The correlated case is obtained with a Gaussian spatial correlation with $\sigma=1$ as explained in the main text.}
    \label{fig:slab_alloy}
\end{figure}

\begin{figure}[htb!]
    \includegraphics[width=\columnwidth]{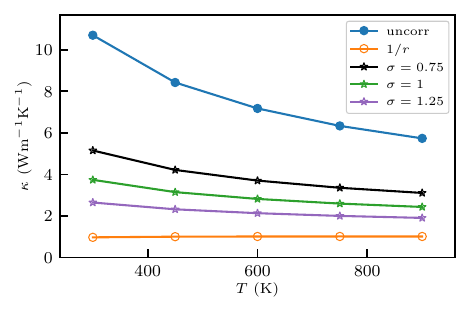}
    \includegraphics[width=\columnwidth]{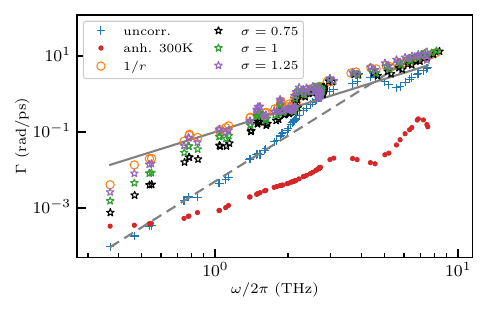}

    \caption{ Upper panel: lattice thermal conductivity for different kinds of colored disorder, as a function of temperature.
     Stars and empty circles indicate respectively a Gaussian and $1/r$ spatial correlation function. $\sigma$ is a dimensionless parameter proportional to the width of the Gaussian, as specified in the text. Lower panel: corresponding mass disorder linewidths of acoustic phonons as a function of frequency. Dashed and continuous gray lines are guidelines for, respectively, a $\omega^4$ and $\omega^2$ behavior. For comparison, we also show with red dots the anharmonic contribution to linewidths at room temperature, and with blue crosses the contribution from disorder scattering for a standard, spatially uncorrelated, alloy. All samples are at $x=0.5$ concentration. }
    \label{fig:corr_scattering}
\end{figure}
\begin{figure}[htb!]
    \includegraphics[width=\columnwidth]{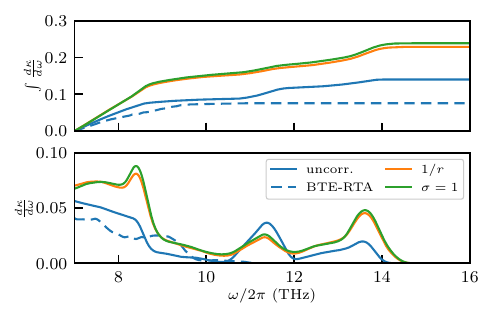}
    \caption{Conductivity accumulation function $\int_{\omega_a}^\omega \frac{d\kappa}{d\omega'}d\omega' $ (top panel) and frequency-resolved differential conductivity $\frac{d\kappa}{d\omega}$ (lower panel), as defined in Ref.~\onlinecite{isaeva2019modeling}, in units of $\mathrm{W m^{-1}K^{-1}}$  and $\mathrm{W m^{-1}K^{-1}ps^{-1}}$, respectively, for different instances of disorder at $T=600\,\mathrm{K}$ and $x=0.5$. The lower limit for the integral in the accumulation function is set to $\omega_a/2\pi=7\mathrm{~THz}$, to neglect differences in the acoustic frequency range due to the different nature of disorder. The continuous lines are QHGK calculations, while the dashed blue line is the BTE-RTA result for the uncorrelated case.}
    \label{fig:kappa_cumulative}
\end{figure}

In the pursuit of reducing the predominant contribution of acoustic modes and lowering $\kappa$ in SiGe thermoelectrics,
several approaches have been proposed based on the principle of enhancing acoustic phonon scattering. 
From concentration graded superlattices~\cite{ferrando-villalba_beating_2020,ferrando-villalba_tailoring_2015} to nanopores \cite{lee_effects_2010} and nanograins~\cite{wang_enhanced_2008,liu2009high,wang_enhanced_2008, joshi_enhanced_2008, zhu_increased_2009, pernot_precise_2010, bathula2012enhanced}, the overall qualitative idea is to produce a disorder-induced scattering of large-wavelength acoustic vibrations, which would otherwise propagate virtually like in a homogeneous medium. Quantitatively, the main limitation in reducing the acoustic contribution through harmonic disorder is represented by Rayleigh's $\omega^4$ scattering, which vanishes so quickly for vanishing frequencies that it cannot even guarantee a finite bulk thermal conductivity in a purely harmonic glass~\cite{chauduri_heat_2010, fiorentino2023unearthing}. 
Under very general assumptions, any harmonic scattering follows Rayleigh's scaling in the hydrodynamic limit~\cite{ganter2010rayleigh}.
However, nontrivial behavior is determined by the spatial correlations of elastic deformations.
Essentially, as long as their correlation length is finite, acoustic vibrations with a wavelength larger than some characteristic length are scattered like the vibrational modes of a chain of equally spaced atoms connected by springs with uncorrelated random spring constants yielding $\Gamma\propto \omega^4$~\cite{allen1998evolution}. At higher frequencies, random media theory predicts a $\omega^4 \to \omega^2$ crossover of the harmonic scattering linewidths~\cite{mizuno2020sound,schirmacher2024nature} which is observed both experimentally~\cite{baldi2011elastic,masciovecchio2006evidence} and numerically~\cite{fiorentino2023unearthing} in amorphous materials such as $a$-SiO$_2$.
If the correlation length increases, the crossover shifts to lower frequencies, thus enhancing the overall scattering strength due to harmonic disorder (see Refs.~\onlinecite{thebaud2023breaking,chaney2021tuneable} and Appendix~\ref{app:perturbative gamma}). 

We investigate two cases of correlated mass distributions, 
\begin{equation}\label{eq:mass correlation function}
    C(r) \propto \frac{1}{N_\mathrm{atoms}}\sum_{I,J=1}^{N_\mathrm{atoms}}\delta M(\mathbf R_J)\delta M(\mathbf R_I)\delta(r-R_{IJ}),
\end{equation}
where $R_{IJ}=|\mathbf R_I-\mathbf R_J|$. One is an infinite-length power-law $C(r) \sim r^{-1}$, and the other a finite-length Gaussian correlation $C(r) \propto e^{-r^2/(2\sigma^2l_0^2)}$, where $\sigma$ is a dimensionless parameter and  $l_0$ is the cubic root of the volume $l_0=V^{1/3}$. First, we generate different mass configurations at $x = 0.5$ for each correlation function using the algorithm described in Appendix~\ref{app:generation disorder}, exemplified in Fig.~\ref{fig:slab_alloy} where a clustering effect can be observed as the result of correlations. Our procedure slightly differs from the commonly used algorithm for generating spatially correlated disorder~\cite{thebaud2023breaking} to reduce noise due to the discreteness of mass values. 

Colored disorder significantly reduces the lattice thermal conductivity, as shown in the upper panel of Fig.~\ref{fig:corr_scattering}. The longer the correlation length the more effective the thermal conductivity reduction. This reduction relies on the enhanced dampening of the acoustic modes, lower panel Fig.~\ref{fig:corr_scattering}, which is not compensated by the slight increase of the diffusive contribution shown in Fig.~\ref{fig:kappa_cumulative}. Apart from being overall stronger, the colored scattering presents qualitative differences from the uncorrelated case.
%
%
Indeed, for frequencies below approximately $2\,\mathrm{THz}$, a crossover from $\omega^4$ to $\omega^2$ for the harmonic scattering is observed in the Gaussian case, with the turning-point frequency increasing monotonically with $\sigma$. To quantitatively capture the crossover it is essential to compute the Debye-like contribution correctly. Therefore we use the empirical function proposed in Ref.~\onlinecite{baldi2011elastic}:
\begin{align}\label{eq:gamma crossover}
    \Gamma^\circ(\omega) = C \omega^2 [1 +(\omega_{c}/\omega)^{2\delta}]^{-1/\delta},
\end{align}
where the fitting parameters $C$ and $\omega_c$ are a prefactor and the crossover angular frequency, respectively. The fixed parameter $\delta=1.5$ determines the sharpness of the transition. Although the $1/r$ correlation theoretically should maintain an $\omega^2$ scaling in the vanishing-frequency limit~\cite{thebaud2023breaking}, it also appears to exhibit a crossover: this is justified by the implicit cutoff due to the imposed PBCs, which prevents the range of correlations to exceed the system's size (see Appendix~\ref{app:perturbative gamma}). Therefore the harmonic scattering has been extrapolated as $\sim \omega^2$, ignoring the crossover which in this case is a finite-size effect.

\subsection{Thermoelectric Figure of Merit}

While the lowest value of thermal conductivity is reached for the $1/r$ correlation, we focus the rest of our analysis on the Gaussian case, whose scattering can be captured without finite-size effects and whose technological implementation would require control only at the nanoscale, rather than at all the scales. Notably, a Gaussian correlation with $\sigma=1.25$ already reduces the thermal conductivity to approximately $2.4\,\mathrm{Wm^{-1}K^{-1}}$ at room temperature, which is about $4.5$ times smaller than the corresponding value for an uncorrelated alloy.

\begin{figure}[htb!]
    \includegraphics[width=\columnwidth]{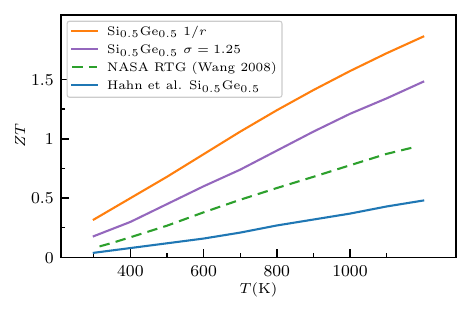}
    \caption{Predicted enhancement of the TE figure of merit of Si$_{0.5}$Ge$_{0.5}$ due to the thermal conductivity reduction from correlated disorder, compared to the figure of merit of NASA RTG from Ref.~\onlinecite{wang_enhanced_2008} and numerical data from Ref.~\onlinecite{hahn2021intrinsic}. The carrier concentration is ${n=10^{20}\,\mathrm{cm^{-3}}}$ for the solid lines and ${n=2\times 10^{20}\,\mathrm{cm^{-3}}}$ for the dashed one.}
    \label{fig:ZT}
\end{figure}

The large thermal conductivity reduction from correlated mass disorder may improve the SiGe TE figure of merit, extending the operational range of SiGe-based TE devices, so far limited to high temperatures, $T>1000$~K, mostly in radioisotope thermoelectric generators (RTGs) powering NASA space probes. RTGs consist of both p- and n-doped polycrystalline SiGe with carriers concentration $n\sim 10^{20}\,\mathrm{cm^{-3}}$ and thermal conductivity lower than bulk crystalline SiGe $\kappa=4.5$-$4.0\,\mathrm{Wm^{-1}K^{-1}}$~\cite{wang_enhanced_2008}. The $ZT$ of state-of-the-art n-doped RTG is shown in Fig.~\ref{fig:ZT}. 
The intrinsic electronic transport coefficients of SiGe were computed by DFT-eVCA and the semiclassical Boltzmann transport equation in Ref.~\onlinecite{hahn2021intrinsic} as a function of alloy composition, temperature, and carrier concentration. We calculated $ZT$ for Si$_{0.5}$Ge$_{0.5}$ with correlated disorder $\sigma=1.25$, using $\sigma_\mathrm{el}$ and $S$ for n-doped SiGe with $n\sim 10^{20}\,\mathrm{cm^{-3}}$ from~\cite{hahn2021intrinsic}.
To obtain $\kappa_\mathrm{el}$ we use Wiedemann-Franz law with an experimentally determined Lorenz number of $2.14\times10^{-8}\,\mathrm{V^2K^{-2}}$. Fig.~\ref{fig:ZT} shows an enhancement of the intrinsic $ZT$ of single-crystal SiGe with uncorrelated disorder from 4-fold at low temperature to 3-fold at high temperature, and a 1.5-fold improvement over the n-doped NASA RTG at all temperatures. 
Repeating the same calculation for the $1/r$ correlated mass disorder yields a theoretical maximum $ZT$ of 1.9 at $T=1200\,\mathrm{K}$. $ZT$ reaches the technologically critical value of 1 at $T=700\,$K, thus making SiGe a potentially efficient TE material at moderate temperatures.

\section{Conclusions}

In summary, we have computed the thermal conductivity of Si$_{1_x}$Ge$_x$ alloy as a function of concentration and temperature using a non-perturbative approach to treat mass disorder. 
Whereas our predictions for the thermal conductivity in the uncorrelated case are similar to the standard perturbative BTE method~\cite{garg2011role}, our approach based on QHGK~\cite{isaeva2019modeling} and hydrodynamic long-wavelength extrapolation~\cite{fiorentino2023hydrodynamic} is more general and allows us to treat systems with correlated disorder. Our calculations predict that the lattice thermal conductivity of Si$_{1_x}$Ge$_x$ ($x=0.5$) may be reduced by a factor $4.5$ at room temperature by engineering correlated disorder. 
The resulting thermal conductivity is similar to that obtained by hot-pressing alloyed nanopowders, and lower than that of state-of-the-art NASA  radio-isotope thermoelectric generators (RTGs) made of polycrystalline SiGe.  
We have demonstrated that the lattice thermal conductivity reduction from correlated mass disorder leads to a factor 1.5 enhancement over the RTGs $ZT$ across the 300-1100~K temperature range and a 4-fold enhancement over the $ZT$ of single crystalline Si$_{0.5}$Ge$_{0.5}$ with uncorrelated disorder.

Whereas manufacturing alloys with correlated compositional disorder may be challenging, SiGe alloy thin films with controlled concentration profiles in the growth direction can be grown epitaxially~\cite{ferrando-villalba_tailoring_2015} and exhibit substantial thermal conductivity reduction~\cite{ferrando-villalba_beating_2020}, indicating that long-range correlations are effective at suppressing heat transport.
 


\medskip

The codes that support the relevant results within this paper are publicly available from the respective developers' repositories. Analysis scripts to compute VDSF and perform the hydrodynamic extrapolation are available on GitHub~\cite{Fiorentino_hydro_glasses_2023} and on the Materials Cloud platform~\cite{talirz2020materials}. See DOI:[to be included when available].

\begin{acknowledgments}
The authors are grateful to Enrico Drigo, Florian Pabst, and Giacomo Tenti for fruitful discussions. We are also grateful to Luciano Colombo and Claudio Melis for providing raw data on first-principles calculations of the transport coefficients of SiGe. This work was partially supported by the European Commission through the \textsc{MaX} Centre of Excellence for supercomputing applications (grant number 101093374), by the Italian MUR, through the PRIN project ARES (grant number 2022W2BPCK), and by the Italian National Centre for HPC, Big Data, and Quantum Computing (grant number CN00000013), funded through the \emph{Next generation EU} initiative. DD acknowledges support from the DARPA Thermonat program (Agreement No. HR00112390126). 
\end{acknowledgments}

\newpage
\appendix
\section{Computational details}\label{app:computational}
Second- and third-order interatomic force constants (IFC) are obtained from standard  \cite{Baroni1987a,*Giannozzi1991a,*baroni2001phonons} and third-order \cite{Gonze1989,*Debernardi1994,*Debernardi1995} DFPT, using the \texttt{pw.x/ph.x} code in the QE distribution~\cite{giannozzi2009quantum,*giannozzi2017advanced,*giannozzi2020quantum} and \texttt{D3Q}~\cite{paulatto2013anharmonic}, respectively.
%
Starting with norm-conserving pseudopotentials based on the approach of von Barth and Car~\cite{dal1993ab} for both silicon and germanium, the virtual crystal pseudopotentials for intermediate concentrations are generated using a dedicated tool within the QE distribution~\cite{giannozzi2009quantum}. For all concentrations, the self-consistent calculations on the relaxed virtual diamond crystals are performed on a $[6,6,6]$ Monkhorst-Pack~\cite{monkhorst1976special} mesh, with an energy cutoff of $24 , \mathrm{Ry}$ for the plane-wave expansion and a convergence threshold of $10^{-12}$. The second- and third-order IFCs are then computed with a threshold of $10^{-16}$, using a $[7,7,7]$ and $[5,5,5]$ supercell, respectively. The ab initio calculations are conducted for the following set of concentrations: $x = 0, 0.1, 0.3, 0.5, 0.7, 0.9, 1$, while the lattice parameters and IFCs for any other concentration are obtained by linear interpolation of the two nearest ab-initio concentrations.

We use these IFCs for the vVCA calculations of anharmonic linewidths and thermal conductivity, as well as for the generation of disordered alloys in real space. The thermal conductivity calculations, both in k-space and real space, are performed using the $\kappa$ALD$o$ program~\cite{barbalinardo2020efficient}, while the VDSF is computed through the "hydro-glass" code available on GitHub~\cite{Fiorentino_hydro_glasses_2023}. For the crystalline case, we use a dense $q$-mesh $[28,28,28]$. For the disordered alloys, the thermal conductivity and VDSF calculations are performed on sizes corresponding to $[18,18,18]$ and $[28,28,28]$ supercells, respectively. 
To take into account the stochastic noise of the generation of the disordered alloys, for each size the results are averaged over $4$ samples.

\section{Haydock's recursion technique}\label{app:haydock}

Haydock's recursion technique is an iterative method, based on Lanczos' iterative algorithm, to efficiently estimate the imaginary part of the diagonal elements of the vibrational Green's functions
 and consequently the spectral functions $S_{\phi}^\circ(\omega)$:
 \begin{multline}
    \lim_{\eta\to0} \Im\expval{\left((\omega+i\eta)^2 -D\right)^{-1}}{\phi}=\\ \frac{\pi}{2\abs{\omega}}\left[S_{\phi}^\circ(\omega)+S_{\phi}^\circ(-\omega) \right],
\end{multline}
where $D$ is the dynamical matrix and $\phi$ a generic vector. Haydock's method\cite{haydock1980recursive,vast2000effects,fiorentino2023unearthing} allows one to compute the continued fraction expansion of the above expression, whose coefficients are evaluated by a recursive Lanczos chain. 
The advantages of Haydock's technique are three-fold. Firstly, the method is known to be numerically robust, despite the well-known instabilities of the Lanczos tridiagonalization scheme~\cite{paige1980accuracy}, and approximately $r=100$ recursion steps were sufficient in our case to retrieve the acoustic phonon linewidth, in agreement with Ref.~\onlinecite{vast2000effects,fiorentino2023unearthing}. Secondly, once the coefficients are computed the estimation of $S_{\phi}(\omega)$ for any $\omega$ is inexpensive. Moreover, perturbative anharmonic effects can be added effortlessly by imposing $\eta=\gamma(\omega)$, where $\gamma(\omega)$  is a fit of the anharmonic linewidths~\cite{fiorentino2023hydrodynamic}. However, we found that for the $\Gamma_{\mathbf q b}$ estimation this method would give no significant difference from the Matthiessen rule~\cite{matthiessen1862influence} $\Gamma_{\mathbf q b}=\Gamma^\circ_{\mathbf q b}+\gamma(\omega)$, where $\Gamma^\circ_{\mathbf q b}$ is the linewidth extracted by the harmonic $S^\circ_{\mathbf q b} $. Finally and most importantly, Haydock's technique scales as $O(rN)$ if $D$ is sparse in the chosen basis. This is a major speed-up from the direct computation of $S_{\phi}^\circ$, Eq.~\ref{eq:harmonic_VDSF}, which involves the diagonalization of $D$ and its cubic scaling. 

\section{Raman spectrum}\label{app:raman}

For a given polarization of the incident and scattered light, the nonresonant Stokes Raman spectra are derived from the spectral function, $S_{\phi}(\omega)$, computed via the Haydock algorithm~\cite{vast2000effects,haydock1980recursive}:

\begin{align*}
    S^{R}(\omega)\propto\frac{n(\omega,T)+1}{\omega}S_\phi(\omega),
\end{align*}
where, $(n(\omega,T)+1)$ takes into account the effect of the Bose-Einstein distribution~\cite{lazzeri2003first}. The vector $\phi$ at the $I$-th position and $\lambda$ direction is defined as 

\begin{align*}
    \phi_{I\lambda}=\frac{1}{\sqrt{M_I}}\sum_{\alpha\beta}e^{in}_\alpha e^{out}_\beta\chi_{I\alpha\beta\lambda},
\end{align*}
where $e^{\mathrm{in}}_\alpha,e^{\mathrm{out}}_\beta$ represent the polarizations of the incident and scattered light, respectively, and $\chi_{I\alpha\beta\lambda}$ is the Raman tensor. Under the eVCA, the Raman tensor is periodic and can be easily computed for the unit cell of the virtual crystal~\cite{vast2000effects}. Finally, The unpolarized Raman spectrum is obtained from the rotation invariants using the powder formula~\cite{prosandeev2005first}, with a computational cost of a few polarized spectra.
\section{Correlated mass disorder}\label{app:generation disorder}
We employ the algorithm from Ref.~\onlinecite{thebaud2023breaking} to create a spatially correlated mass distribution, followed by a Monte Carlo minimization to refine the solution and reduce the errors due to our (very) discrete set of masses: $m_{\Ge}$ and $m_\Si$. We focused on the symmetrical case with Germanium concentration $x=0.5$.

Given a desired correlation function $C(r)$, the goal is to build a residual mass distribution ${\delta M(r)=M(r)-N_\mathrm{atoms}^{-1}\sum_I M_I}$ according to Eq.~\ref{eq:mass correlation function}.

This can be accomplished thanks to the convolution properties of the Fourier Transform (FT), computed with the Fast Fourier Transform algorithm. In fact, the mass distribution must satisfy the following equation:
 \begin{align}
     \delta \tilde M(\mathbf{q})=\sqrt{|\tilde C(\mathbf q)|} e^{i\varphi(\mathbf q)}
 \end{align}
 where $\varphi(\mathbf q)$ is an arbitrary phase. Stochastically equivalent configurations can be generated By extracting this arbitrary phase from a uniform distribution $[0,2\pi]$, with the additional requirement of $\varphi(q)=-\varphi(-q)$ to guarantee a real $\delta M(r)$ for a crystal with inversion symmetry. Since we are interested in correlation on scales larger than the unit cell, $\sim 5 \AA$, we operated on the lattice, therefore treating the masses of the two atoms of the unit cell as a whole. 
 
\begin{figure}[htb!]
    \includegraphics[width=\columnwidth]{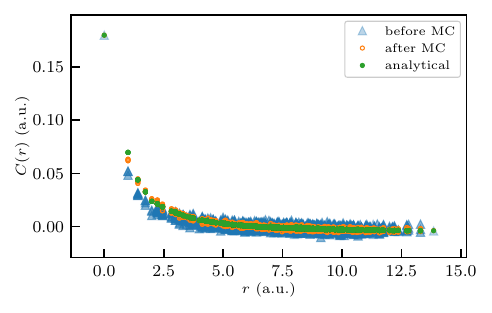}
    \caption{ Example of mass correlation distribution before and after the Monte Carlo refining procedure. All the quantities are computed on a cubic lattice with lattice parameter $a=1$ and supercell $[16,16,16]$ for a desired correlation function $C(r)\propto \frac{e^{-\epsilon r}}{r}$, where $\epsilon=10^{-4}$. }
    \label{fig:Cr_MC}
\end{figure}

By Inverse FT (IFT), we retrieve $\delta M(r)$. Then, we discretize it :
\begin{equation*}
    \delta M^D(r_I)\propto\mathrm{sign}(\delta M(r))
\end{equation*}
while conserving its variance and ensuring a zero average. Assigning $m_{\Ge}$ for positive values and $m_{\Si}$ for negative ones, or vice versa, would conclude the algorithm from Ref.~\onlinecite{thebaud2023breaking}. Without a finer grid of mass values, e.g. Ref.~\onlinecite{thebaud2023breaking} used $5$  values, the discretization introduces noise. In order to find the optimal $M^D$ we implemented a Monte Carlo algorithm to minimize the loss function:
\begin{equation*}
    \mathcal{Loss}(t)=\sum_{q\neq 0} |\Tilde C (q)- |\Tilde{M}^D(q,t)|^2|
\end{equation*}
where $t$ indicates the Monte Carlo step. Each Monte Carlo move consists of switching two elements $i,j$ of $M^D(r_I)\rightleftharpoons M^D(r_J)$. $q=(0,0,0)$ is excluded by the loss function since $C(q=(0,0,0))\propto \sum_{ij}M_IM_J$ cannot be modified by the Monte Carlo move and it is determined by the average of $\delta M^D(r)$, initially imposed to zero. An example of the effect of this Monte Carlo refining is shown in Fig.~\ref{fig:Cr_MC}.
 
We employed this algorithm with a short-ranged and a long-ranged correlation function, respectively $e^{-r^2/(2\sigma^2 l_0^2)}$ and $e^{-\epsilon r}/r$. Being $a$ the lattice parameter of the diamond crystal, the parameters indicate respectively the cubic root of the volume $l_0=V^{1/3}=(a^3/4)^{1/3}$, an adimensional parameter $\sigma$ to determine the cutoff of the Gaussian, and finally $\epsilon$ is a regularization parameter order of magnitudes smaller than the inverse of our largest side $L$ to avoid the non-analytical part of the FT of $1/r$  without altering the $\sim 1/r$ behavior for $r<L$ excessively. 

\section{Details on the hydrodynamic extrapolation}\label{app:hydro_ext}

\begin{figure}[htb!]
    \includegraphics[width=\columnwidth]{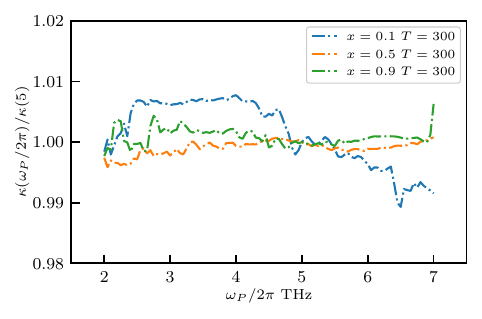}
    \caption{ Bulk lattice thermal conductivity as a function of $\omega_{\mathrm{P}}$ divided by its value for $\omega_{\mathrm{P}}/2\pi=5\mathrm{THz}$, for different concentrations at room temperature. }
    \label{fig:test_omega_p}
\end{figure}

In the main text and Ref.~\onlinecite{fiorentino2023hydrodynamic} the choice of $\omega_\mathrm{P}$ is discussed as the frequency separator between the acoustic crystalline-like contribution and the ``diffusive'' contribution computed on the disordered system. While $\omega_\mathrm{P}$ reminds and is related to the crossover frequency between propagons and diffusons, determined with the Ioffe-Regel criterion~\cite{ioffe1960progress}, it is usually smaller and it allows more versatility of choice. The choice of $\omega_\mathrm{P}$ is arbitrary as long as it happens in a region where the real-space disordered contribution is well-converged and the conditions for the acoustic basis are satisfied, i.e. the VDSF shows well-separated longitudinal and transverse acoustic bands. This is exemplified in Fig.~\ref{fig:test_omega_p}, where it can be noticed the thermal conductivity as a function of $\omega_\mathrm{P}$, relative to the result for $\omega_\mathrm{P}/2\pi=5\mathrm{~THz}$. Indeed, between $2$ and $7\,\mathrm{THz}$, the relative variation for three different concentrations is less than $1\%$.
\section{Effect of mass disorder on anharmonicity}\label{app:mass_effect_anh}
\begin{figure}[htb!]
    \includegraphics[width=\columnwidth]{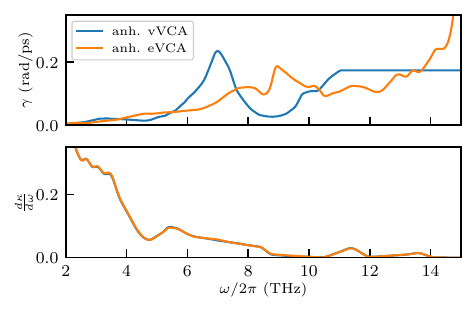}
    \caption{Top panel: frequency-interpolated anharmonic linewidths in the eVCA and vVCA for $x=0.5$. Lower panel: corresponding frequency-resolved thermal conductivity in units of $\mathrm{Wm^{-1}K^{-1}ps^{-1}}$.   }
    \label{fig:test_gamma_p}
\end{figure}

Harmonic disorder can significantly affect anharmonic scattering. Indeed, the anharmonic linewidths computed with third-order anharmonicity and FGR\cite{fabian1996anharmonic} depend on both the eigenvectors and eigenvalues of the (disordered) dynamical matrix. However, in Fig.~\ref{fig:test_gamma_p} we show that despite the considerable difference between the virtual crystal anharmonic linewidths and the ones of a mass-disordered alloy, the thermal conductivity is not affected.

In the top panel Fig.~\ref{fig:test_gamma_p} we compare the virtual crystal's linewidths with the ones computed on a $250$ atoms (uncorrelated) mass-disordered alloy, both at $x=0.5$. For the disordered alloy, the inter-atomic force constants are still computed in the eVCA approximation, using the Hiphive code\cite{eriksson2019hiphive}, but with the eVCA, disordered, eigenvectors.

For both systems, we interpolated the anharmonic linewidths in frequency with a spline. Since the maximum frequency of the virtual crystal is lower than the one of the disordered alloy, we extrapolated with a constant. The two splines tend to overlap at lower frequencies while they differ significantly at higher frequencies. However, such a difference does not affect the frequency-resolved thermal conductivity plotted in the lower panel.  As a rule of thumb, as long as the anharmonic difference is located in a frequency region dominated by disorder, g.e. the diffusive region, its effect on thermal conductivity is negligible. Therefore, our choice, shared by Ref.~\onlinecite{thebaud2020success,garg2011role}, of using the virtual crystal linewidths does not reduce the accuracy of our calculations.

\section{Scattering rate scaling in finite systems}\label{app:perturbative gamma}

Calculations on atomistic systems in PBCs are characterized by an implicit cutoff distance given by the supercell size, $L$. Assuming a long-range mass perturbation correlation function $C(r) \propto r^{-1}$ for an ideal system, the respective correlation function for an actual calculation is of the form $C(r) \propto r^{-1} \Theta(L-r)$, where $\Theta$ is the Heaviside step function. In the Debye limit, the perturbative scattering rate can be expressed in terms of the Fourier transform of $C(r)$, $\tilde{C}(q)$, as~\cite{tamura1983isotope, thebaud2023breaking} 
\begin{align}
    \Gamma_{\mathbf{k}} \propto \omega_\mathbf{k}^3 \int \tilde{C}(q)  \frac{q^2}{2k^2} \delta(c^2(\mathbf{k} + \mathbf{q})^2 - c^2 k^2) \dd^3{q},
\end{align}
where $c$ is the speed of sound. Given the assumed form of $C(r)$, its Fourier transform is
\begin{align}
    \begin{split}
        \tilde{C}(q) &= \frac{1}{(2\pi L)^3} \int \frac{1}{r} \Theta(L-r) e^{i \mathbf{q} \cdot \mathbf{r}} \dd^3 r \\
        &= \frac{2-2 \cos(L q)}{(2\pi L)^3 q^2},
    \end{split}
\end{align}
which in turn yields
\begin{equation}\label{eq:gamma long-range cutoff}
    \Gamma_{\mathbf{k}} \propto \frac{1}{4 L^5 \pi^3} (1 - 2 k^2 L^2 - \cos(2 k L) - 2 k L \sin(2 k L)).
\end{equation}
For any finite $L$, the dominant term for low $k$ is $k^4$. For infinite $L$, the dominant contribution is $k^2$~\cite{thebaud2023breaking}.
The crossover wavenumber can be found considering a dimensionless version of~\eqref{eq:gamma long-range cutoff}, 
\begin{align}
    \Gamma_{\mathbf{k}} \propto (1 + 2 \xi^2) \left(1 - \frac{2 \xi \sin(2\xi) + \cos(2 \xi)}{1 + 2 \xi^2}\right),
\end{align}
where $\xi=kL$, and equating the two asymptotic limits,
\begin{align*}
    2 \xi^4, & \;\text{for } \xi \to 0, \\
    1 + 2 \xi^2, & \; \text{for } \xi \to \infty,
\end{align*}
yielding $k_{\mathrm{cross}} = \frac{1}{L} \sqrt{\frac{1+\sqrt{3}}{2}}$.

\newpage
\bibliography{main}

\end{document}